\documentclass[twocolumn,superscriptaddress,showpacs,amsmath,amssymb,aps, prb]{revtex4-2}

\usepackage{graphicx}
\usepackage{dcolumn}
\usepackage{color}
\usepackage{bm}

\newcommand{\tG}{\tilde{G}}
\newcommand{\trho}{\tilde{\rho}}

\usepackage[whole]{bxcjkjatype} 

\begin{document}


\title{Robust analytic continuation using sparse modeling approach imposed by semi-positive definiteness for multi-orbital systems}

\author{Yuichi Motoyama}
\affiliation{Institute for Solid State Physics, University of Tokyo, Chiba 277-83581, Japan}

\author{Hiroshi Shinaoka}
\affiliation{Department of Physics, Saitama University, Saitama 338-8570, Japan}

\author{Junya Otsuki}
\affiliation{Research Institute for Interdisciplinary Science, Okayama University, Okayama 700-8530, Japan}

\author{Kazuyoshi Yoshimi}
\affiliation{Institute for Solid State Physics, University of Tokyo, Chiba 277-8581, Japan}

\date{\today}

\begin{abstract}
Analytic continuation (AC) from imaginary-time Green's function to spectral function is essential in the numerical analysis of dynamical properties in quantum many-body systems.
However, this process faces a fundamental challenge: it is an ill-posed problem, leading to unstable spectra against the noise in the Green's function.
This instability is further complicated in multi-orbital systems with hybridization between spin-orbitals, where off-diagonal Green's functions yield a spectral matrix with off-diagonal elements, necessitating the matrix's semi-positive definiteness to satisfy the causality.
We propose an advanced AC method using sparse modeling for multi-orbital systems, which reduces the effect of noise and ensures the matrix's semi-positive definiteness.
We demonstrate the effectiveness of this approach by contrasting it with the conventional sparse modeling method, focusing on handling Green's functions with off-diagonal elements, thereby demonstrating our proposed method's enhanced stability and precision.
\end{abstract}

\maketitle


\section{\label{sec:intro}Introduction}

The imaginary-time representation is a foundation in calculating quantum many-body systems at finite temperatures in theory and practice.
Analytical methods leverage this representation to calculate thermal expectation values for a spectrum of static and dynamic metrics via the imaginary-time Green's function and diagrammatic approaches~\cite{AbrikosovGD1963}.
When calculating the spectral function $\rho(\omega)$ of quantum many-body systems using numerical methods based on imaginary-time path integral, such as quantum Monte Carlo methods~\cite{GullMLRTW2011,GubernatisKW2016}, it is necessary first to obtain the temperature Green's function $G(\tau)$ and solve the integral equation, $G(\tau)=\int d\omega e^{-\tau \omega}/(1\pm e^{-\beta \omega})\rho(\omega)\equiv \int d\omega K_{\pm}(\tau, \omega) \rho(\omega)$,
where $\beta$ is the inverse temperature.
The kernel function $K$ exponentially decays for the frequency $\omega$.
Hence, the obtained spectral function is unstable against noise in the temperature Green's function (e.g., statistical errors of QMC), especially in the high-frequency region. 

To solve this problem, various methods have been proposed;
the Pad\'{e} approximation method~\cite{VidbergS1977, Schtt2016, Han2017, Kiss2019,Weh2020},
the maximum entropy method~\cite{SilverSG1990, JarrelG1996, Dirks2010, Reymbaut2015, Bergeron2016, Kraberger2017, Sim2018, Yang2024arxiv},
the stochastic method~\cite{Sandvik1998, MishchenkoPSS2000, Beach2004, FuchsPJ2010, Sandvik2016, Bao2016, ShaoQCCMS2017},
the Nevanlinna or the Carath\'{e}odory formalism~\cite{Fei2021, Fei2021b, Nogaki2023},
the pole representation method~\cite{Huang2023, Huang2023b, Zhang2023-aa},
and the regression using neural networks~\cite{YoonSH2018, FournierWYW2020, KadesPRSUWWZ2020, XieBMW2021}.
We have recently developed another robust method based on an intermediate representation basis\cite{ShinaokaOOY2017} using the singular value decomposition of the integration kernel and the sparse modeling (SpM)\cite{OtsukiOSY2017, Yoshimi:2019bt}, as well as a method combining SpM and Pad\'{e} approximation (SpM-Pad\'{e})\cite{Motoyama2022}.
In these studies, we have considered only diagonal Green's functions where the creation and the annihilation operators act on the same orbitals. 

When dealing with actual materials, systems often have multiple orbitals.
In such systems, off-diagonal Green's functions appear, and then the spectral function becomes a matrix with off-diagonal elements.
In this case, the semi-positive definiteness (SPD) of the matrix is required to satisfy the causality of the Green's function.
However, if each matrix component is continued separately, the semi-positive definiteness may be violated. 
Therefore, to obtain physically valid results, it is necessary to perform the analytic continuation of the entire matrix at once while satisfying the semi-positive definiteness.
Some numerical AC methods have been extended to matrix-valued functions~\cite{Kraberger2017, Sim2018, Yang2024arxiv, Fei2021b, Huang2023b}.

In this work, we extend the SpM analytic continuation method for multi-orbital systems that ensures the semi-positive definiteness of the matrix. 
Using artificial data, we demonstrate the efficiency of the new method by investigating the difference between the conventional SpM method and our new method.
In addition, we investigate how to reduce the computational cost scaling as the cubic of the number of orbitals.
Section II is devoted to a brief review of the SpM analytic continuation method for single orbital systems and the introduction of its extension to multiple orbital systems.
The numerical demonstrations are shown in Section III.
Finally, we summarize the study in Section IV.

\section{\label{sec:methods}Methods}
\subsection{Single orbital SpM-AC}
The problem to be solved is
\begin{equation}
G(\tau) = \int_{-\infty}^\infty \mathrm{d}\omega K(\tau, \omega) \rho(\omega),\label{eq:gtau}
\end{equation}
where
\begin{equation}
K(\tau, \omega) = \begin{cases}
e^{-\tau\omega} / \left(1 + e^{-\beta\omega}\right) & \text{(Fermion)} \\
e^{-\tau\omega} / \left(1 - e^{-\beta\omega}\right) & \text{(Boson)}
\end{cases}.
\end{equation}
To solve Eq.~\eqref{eq:gtau} for $\rho(\omega)$, first, the maximum and minimum frequency cutoff $\omega_\text{max}$ and $\omega_\text{min}$ are introduced,
and the imaginary time and the frequency are discretized to $N_\tau$ and $N_\omega$ points as
\begin{align}
\tau_i &= i\beta / (N_\tau-1) \\
\omega_\alpha &= \alpha \Delta\omega + \omega_\text{min}
\end{align}
where $i$ and $\alpha$ are the indices of discretized time and frequency, respectively,
and $\Delta\omega=(\omega_\text{max}-\omega_\text{min})/(N_{\omega}-1)$. 
By using them, Eq.(\ref{eq:gtau}) is discretized as
\begin{equation}
G_i = \sum_\alpha K_{i\alpha}\rho_\alpha,
\end{equation}
where
$K_{i\alpha} = K(\tau_i, \omega_\alpha)$,
$G_i = G(\tau_i)$,
and $\rho_\alpha = \rho(\omega_\alpha)\Delta\omega$
Next, we perform the singular value decomposition of the kernel matrix $K$ and represent $G$ and $\rho$ in the IR basis~\cite{ShinaokaOOY2017} as
\begin{eqnarray}
K_{i\alpha} &=& \sum_\ell U_{i\ell} S_\ell V^\dagger_{\ell\alpha} \\
\tG_\ell &=& \sum_i U^\dagger_{\ell i} G_i\\
\trho_\ell &=& \sum_i V^\dagger_{\ell\alpha} \rho_\alpha.
\end{eqnarray}
Hereafter, symbols with the tilde mark like $\tG$ and $\trho$ are quantities in the IR basis.
Therefore, the problem was reduced to
\begin{equation}
\tG_\ell = S_\ell \trho_\ell.
\end{equation}
It should be noted that we can truncate the singular values $S_\ell$ to reduce the computational cost.
To suppress the noise remaining in $\tG$, we introduce $L_1$ regularized cost function to be minimized as
\begin{equation}
L(\trho)
=
\frac{1}{2}\sum_\ell\left(\tG_\ell - S_\ell\trho_\ell\right)^2 + \lambda\sum_\ell|\trho_\ell|,
\end{equation}
where the coefficient of $L_1$ term, $\lambda$, is a hyperparameter of the algorithm.
To minimize the cost function with respect to $\trho$, the alternating direction method of multipliers (ADMM) can be used\cite{BoydPCPE2011}.
ADMM is an extension of the Lagrange multiplier method and considers variable $\trho$ in each term independent and adds constraints between them as
\begin{equation}
\begin{split}
L(\trho) &\rightarrow L(\trho^{(1)}, \trho^{(2)}) \\
&=
\frac{1}{2}\sum_\ell\left(\tG_\ell - S_\ell \trho_\ell^{(1)}\right)^2 + \lambda\sum_\ell|\trho_\ell^{(2)}| \\
&+ \sum_\ell h_\ell (\trho_\ell^{(1)} - \trho_\ell^{(2)}) + \mu\sum_\ell \left(\trho_\ell^{(1)} - \trho_\ell^{(2)}\right)^2,
\end{split}
\end{equation}
and optimizes $\trho^{(1)}$ and $\trho^{(2)}$ in an alternating way.
The $L_1$ regularization term suppresses components $\trho_\ell$ with smaller value than some threshold controlled by $\lambda$.
In addition, we can impose constraints on the spectral function by introducing extra terms to the cost function.
For example, the causality requires the non-negativity of the spectral function, which is
represented by adding the following term,
\begin{equation}
L_\text{non-neg}
=
C_\infty \sum_j \Theta(-\rho_j)
=
C_\infty\sum_j\Theta(-\sum_\ell V_{j\ell}\trho_\ell),
\end{equation}
where $\Theta(x)$ is the step function
and $C_\infty$ is a huge positive number.
One of the advantages of the ADMM is that it works when the number of terms increases.

Once the optimal $\tilde{\rho}_\ell$ is obtained, the spectral function $\rho(\omega)$ is calculated by
\begin{equation}
\rho(\omega_\alpha) = \frac{\rho_\alpha}{\Delta\omega} = \frac{1}{\Delta\omega}\sum_\ell V_{\alpha\ell}\tilde{\rho}_\ell.
\end{equation}

\subsection{Multi-orbital SpM-AC}
Hereafter, let us consider systems with multiple orbitals.
The temperature Green's function and the spectral function become matrices with indices of orbitals, and each component of $G$ and $\rho$ is related to each other in the same way as the single-orbital systems;
\begin{equation}
G_{ab}(\tau) = \int_{-\infty}^\infty \mathrm{d}\omega K(\tau,\omega)\rho_{ab}(\omega),
\label{eq:integral_multi}
\end{equation}
where $a$ and $b$ denote the index of the orbitals.
Discretizing the frequency and imaginary time, this equation can be rewritten as
\begin{equation}
G_{ab,i} = \sum_\alpha K_{i\alpha}\rho_{ab,\alpha},
\end{equation}
where $G_{ab,i} = G_{ab}(\tau_i)$, $\rho_{ab,\alpha} = \rho_{ab}(\omega_\alpha)\Delta\omega$. 
While the non-negative condition of spectral function is required by the causality for single-orbital systems,
the SPD condition, in other words, all the eigenvalues are non-negative, is required for multiple orbital systems.

In this paper, we will compare two types of multi-orbital SpM-AC methods with the SPD condition, called the one-shot (OS) method and the self-consistent (SC) method.
In the OS method, each element $\rho_{ab}(\omega)$ is calculated independently by the single orbital SpM-AC method.
Then, SPD is recovered at each frequency by diagonalizing the obtained $\rho_{ab}(\omega_\alpha)$ and truncating negative eigenvalues as follows:
\begin{equation}
\begin{split}
\rho_{ab}(\omega_\alpha) 
&= 
\sum_c U_{ac}(\omega_\alpha) \Lambda_c(\omega_\alpha) U^\dagger_{cb}(\omega_\alpha) \\
&\simeq 
\sum_c U_{ac}(\omega_\alpha) R\left(\Lambda_c(\omega_\alpha)\right) U^\dagger_{cb}(\omega_\alpha) \\
&=
\hat{\rho}_{ab}(\omega_\alpha),
\end{split}
\end{equation}
where $\Lambda_c(\omega_\alpha)$ is the eigenvalues of $\rho_{ab}(\omega_\alpha)$, and $R(x)$ is a ramp function; $R(x) = x$ if $x>0$ and otherwise $R(x)=0$.

In the SC method, on the other hand, all of the matrix elements of the spectral function are simultaneously reconstructed with the constraint on SPD. In this case, the cost function is defined as follows:
\begin{equation}
\begin{split}
L(\tilde{\rho})
=&
\frac{1}{2}\sum_{a,b,\ell}\left(\tilde{G}_{ab,\ell} - S_\ell \tilde{\rho}_{ab,\ell}\right)^2
+
\lambda \sum_{a,b,\ell}|\tilde{\rho}_{ab,\ell}| \\
&+
C_\infty \sum_\alpha\sum_c\Theta(-\Lambda_{c}(\omega_\alpha)).
\end{split}
\label{eq:MOSPM-cost}
\end{equation}
In solving the optimization problem by ADMM, it is required to perform diagonalization of $\rho_{ab}(\omega_i)$ for all the frequency points on every sweep, causing a slowing down.
To overcome it, we reduce the number of frequencies where the SPD condition is imposed; the last term of Eq.~(\ref{eq:MOSPM-cost}) is replaced with
\begin{equation}
C_\infty\sum_{\alpha'=1}^{N_\omega/n}\sum_c \Theta(-\Lambda_c(\omega_{n\alpha'})),
\label{eq:SPD-skipped}
\end{equation}
where $N_\omega$ is the number of frequencies and $n$ is a hyperparameter controlling how many frequencies are skipped.

\begin{figure*}[tb]
\includegraphics[width=1.0\linewidth]{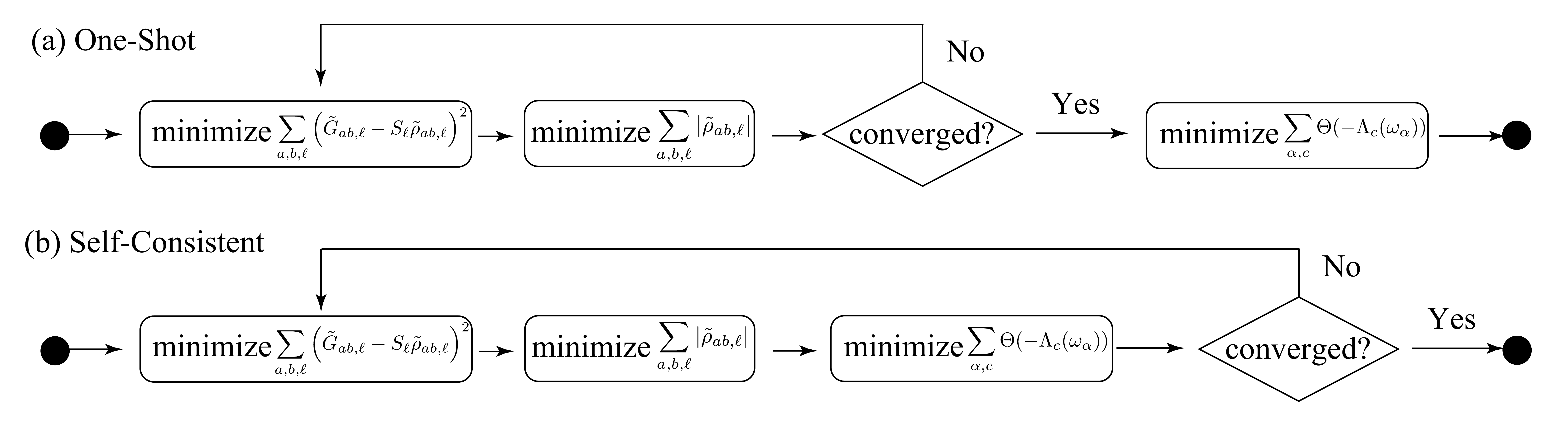}
\caption{
Schematic figure of the optimization processes in (a) the One-Shot method and (b) the Self-Consistent method.
}
\label{fig:loop}
\end{figure*}

To briefly summarize the two methods,
the OS method imposes the SPD condition only once after the self-consistent loop of the ADMM,
whereas the SC method includes the SPD condition in the self-consistent loop (FIG.~\ref{fig:loop}).

\subsection{Optimization of $\lambda$}

\begin{figure}[tb]
\includegraphics[width=1.0\linewidth]{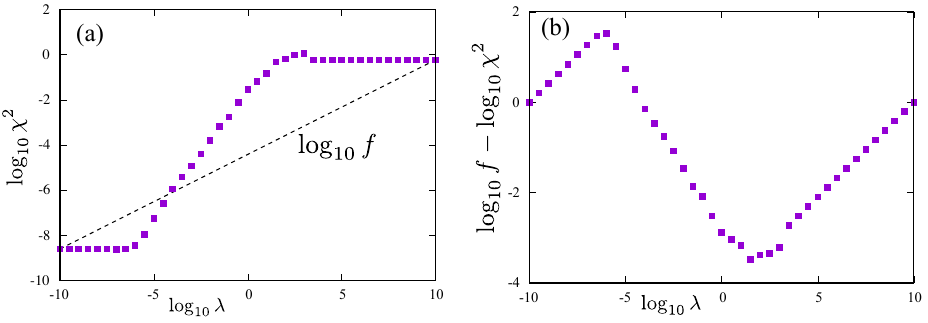}
\caption{
(a) Typical behavior of residues $\chi^2(\lambda)$ in a log-log plot.
(b) The value of $\lambda$ to maximize $\log f(\lambda) - \log \chi^2(\lambda)$ is optimal.
}
\label{fig:elbow}
\end{figure}

The $\lambda$ is a hyperparameter of the algorithm and should be tuned.
In this study, we choose the value of $\lambda$ automatically by using the elbow method~\cite{OtsukiOSY2017}:
First, we optimize $\tilde{\rho}_{ab,\ell}$ with several $\lambda \in [\lambda_\text{min}, \lambda_\text{max}]$ as $\tilde{\rho}_{ab,\ell}^*(\lambda)$.
Then, the residues are calculated as 
\begin{equation}
\chi^2(\lambda) = \frac{1}{2}\sum_{a,b,\ell}\left(\tilde{G}_{ab,\ell} - S_\ell\tilde{\rho}_{ab,\ell}^*(\lambda)\right)^2.
\end{equation}
Figure~\ref{fig:elbow}(a) shows a typical example of a $\chi^2(\lambda)$ curve -- the result of the two-orbital data described in Section~\ref{sec:results-two} obtained by the SC method.
As shown, $\chi^2(\lambda)$ has two plateaus at high and low $\lambda$ regions and one slope connecting them, and the optimal $\lambda$ lies around the lower kink~\cite{OtsukiOSY2017} ($\log_{10}\lambda \sim -6$ in this example).
To detect the position of the lower kink automatically, we maximize $\log f - \log \chi^2$ (see Fig.~\ref{fig:elbow}(b)), where
\begin{equation}
\begin{split}
\log f(\lambda) \equiv&
\frac{\log\chi^2 (\lambda_\text{max}) - \log\chi^2(\lambda_\text{min})}{\log\lambda_\text{max} - \log\lambda_\text{min}} \left(\log\lambda - \log\lambda_\text{min}\right) \\
&+ \log\chi^2(\lambda_\text{min}).
\end{split}
\end{equation}

\section{\label{sec:results}Numerical results}
\subsection{\label{sec:results-two}Two orbital data}
\begin{figure}[tb]
\includegraphics[width=1.0\linewidth]{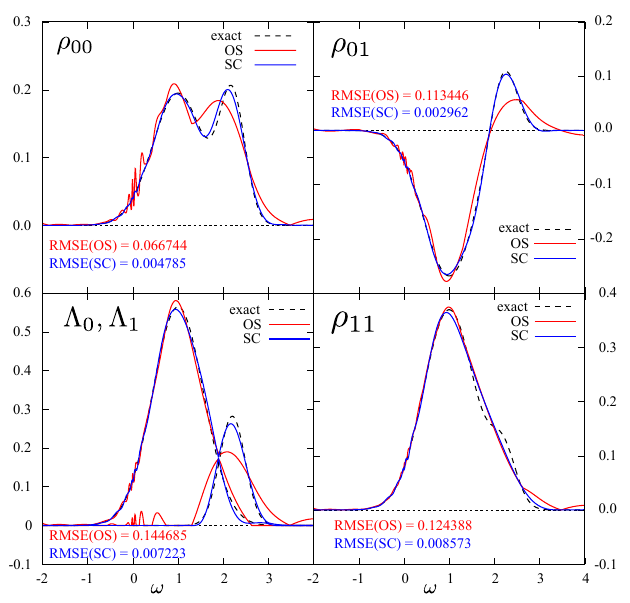}
\caption{
The reconstructed spectral functions $\rho_{ab}(\omega)$ by the OS method (red curve) and the SC method (blue curve) are shown with the exact ones (dashed curve).
The left bottom panel shows the eigenvalues of the spectral function.
}
\label{fig:spectrum}
\end{figure}

For a demonstration of the two methods, we first apply them to an artificial two-orbital spectral function prepared in the following way.
First, the \textit{exact} diagonalized spectral function matrix is defined as 
\begin{equation}
\Lambda^\text{exact}(\omega) =
\begin{pmatrix}
\Lambda^\text{exact}_0(\omega) & 0\\
0 & \Lambda^\text{exact}_1(\omega)
\end{pmatrix}
\end{equation}
with
\begin{equation}
    \Lambda^\text{exact}_0(\omega) = 0.8g(\omega; \mu=1.0, \sigma=0.8)
\end{equation}
and
\begin{equation}
    \Lambda^\text{exact}_1(\omega) = 0.2g(\omega; \mu=2.2, \sigma=0.4),
\end{equation}
where
\begin{equation}
g(\omega; \mu, \sigma) = \sqrt{\frac{1}{\pi\sigma^2}}e^{-\left(\frac{\omega-\mu}{\sigma}\right)^2}.
\end{equation}
Then, the exact spectral function is defined as
\begin{equation}
\rho^\text{exact}(\omega) =
R^\dagger(\theta) \Lambda^\text{exact}(\omega) R(\theta),
\end{equation}
where $R(\theta) = 
\begin{pmatrix}
\cos(\theta) & -\sin(\theta)\\
\sin(\theta) & \cos(\theta)
\end{pmatrix}$
is the rotating matrix and $\theta = 0.3\pi$.
$\rho^\text{exact}_{ab}(\omega)$ and $\Lambda^\text{exact}_a(\omega)$ are shown in FIG.~\ref{fig:spectrum} as the dashed curves.
After constructing $\rho^\text{exact}_{ab}(\omega)$, the temperature Green's function $G_{ab}(\tau)$ is calculated by using Eq.~(\ref{eq:integral_multi}) with $\beta=100$.
At this time, Gaussian noise $N(0,\sigma=10^{-6})$ is added at each imaginary time independently.

FIG.~\ref{fig:spectrum} shows the reconstructed spectral functions $\rho_{ab}(\omega)$ and their eigenvalues $\Lambda_a(\omega)$ by the OS method (red curves) and the SC method (blue curves) with the exact values (dashed curves).
$\text{RMSE}$ means the root mean square error from the exact values,
\begin{equation}
\text{RMSE} = \sqrt{\frac{1}{N_\omega}\sum_\alpha \left(\rho_{ab}(\omega_\alpha) - \rho^{\text{exact}}_{ab}(\omega_\alpha) \right)^2}.
\end{equation}
These RMSEs indicate that SC method gives a closer spectrum to the exact spectrum than the OS method.
Additionally, in the OS method, the truncation of the negative eigenvalue introduces an undesired kink of the spectrum, for example, in $\rho_{00}(\omega \sim 1.2)$.

\begin{figure}[tb]
\includegraphics[width=1.0\linewidth]{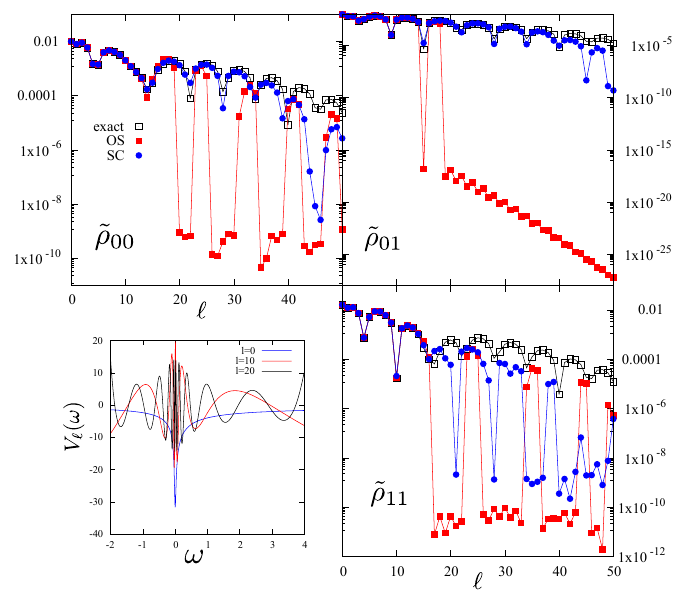}
\caption{
The spectral functions in the IR basis, $|\tilde{\rho}_{ab,\ell}|$, by the OS method (red square) and the SC method (blue circle) are shown with the exact ones (black open square).
The left bottom panel shows the IR basis in the frequency space, $V_\ell(\omega)$ for $\ell = 0$ (blue), $\ell = 10$ (red), and $\ell = 20$ (black).
}
\label{fig:rhol}
\end{figure}

Figure~\ref{fig:rhol} shows the absolute values of the spectral functions represented in the IR basis, $|\tilde{\rho}_{ab,\ell}|$.
While both methods succeed in reconstructing the small $\ell$ components, the OS method fails for the large $\ell$ components.
Because the $V_\ell(\omega)$ with large $\ell$ oscillates rapidly around the origin as shown in the left-bottom panel of Fig.~\ref{fig:rhol}, these deviations from the exact values result in an oscillation of the reconstructed spectral function (see $\rho_{00}$ in FIG.~\ref{fig:spectrum}).
Note that this artifact can be removed by using the SpM-Pad\'{e} method~\cite{Motoyama2022}.

\begin{figure}[tb]
\includegraphics[width=0.9\linewidth]{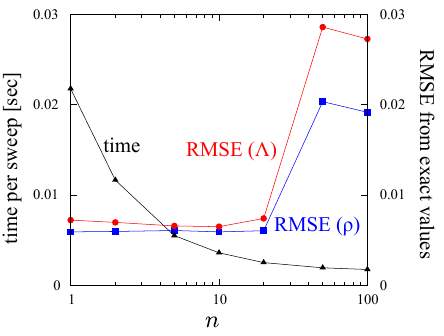}
\caption{
The performance depends on how many frequencies the SPD condition is imposed.
Blue squares and red circles are root mean square errors of the spectrum and eigenvalues (the right axis) and black triangles are time per ADMM sweep in seconds (the left axis).
}
\label{fig:time_freq}
\end{figure}

\begin{figure}[tb]
\includegraphics[width=0.9\linewidth]{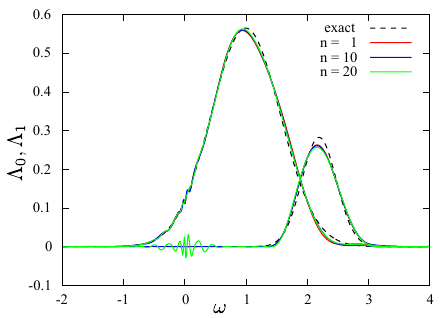}
\caption{
The eigenvalues of the reconstructed spectral functions for $n=1$ (red), $n=10$ (blue), and $n=20$ (green).
}
\label{fig:eigen_freq}
\end{figure}

While the SC method gives a more precise result, the method takes a longer time to evaluate than the other;
for the present demonstration, the SC method took $21.8$ milliseconds while the OS method took $1.6$ milliseconds in one ADMM sweep.
The reduction of the SPD-imposed frequency points [Eq.~(\ref{eq:SPD-skipped})] can speed up the SC method.
Figure~\ref{fig:time_freq} shows how the RMSEs (blue squares for spectrum $\rho$ and red circles for eigenvalues $\Lambda$) and the wall-clock time per ADMM sweep (black triangles) depend on $n$.
As $n$ increases, the RMSEs remain unchanged until $n=10$ but get worse $n>10$.
Indeed, as shown by Fig.~\ref{fig:eigen_freq}, an artificial oscillation of lower eigenvalue appears in the reconstructed eigenvalues with $n=20$, while it does not in that with $n=10$.
On the other hand, six times speedup is achieved when $n=10$ ($3.61$ milliseconds for $n=10$).
In this case, therefore, we can let $n$ be $10$ safely to decrease the computational cost with the accuracy keeping.

\begin{figure}[tb]
\includegraphics[width=1.0\linewidth]{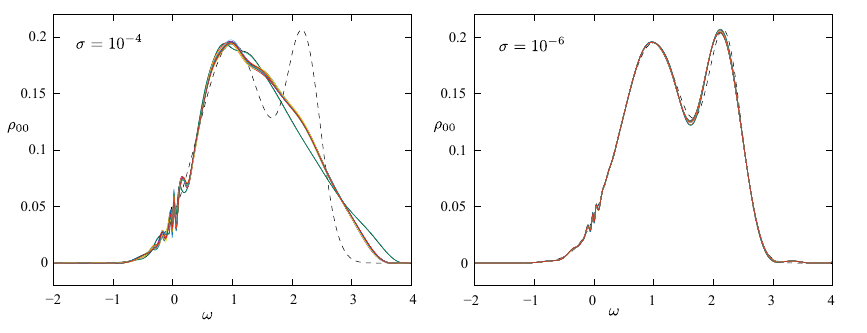}
\caption{
Thirty reconstructed spectral functions $\rho_{00}(\omega)$ of the two-orbital systems from 30 Green's functions with independent Gaussian noise.
The left panel shows the result of a large noise case ($\sigma = 10^{-4}$) and the right panel shows that of a small noise case ($\sigma = 10^{-6}$).
}
\label{fig:spectrum-30samples}
\end{figure}

Finally, to see the robustness of the SC method, we reconstructed spectral functions $\rho_{00}$ from 30 Green's functions with independent Gaussian noise in parallel.
FIG.~\ref{fig:spectrum-30samples} shows the results of the large noise ($\sigma = 10^{-4}$) case (left panel) and the small noise ($\sigma = 10^{-6}$) case (right panel); one curve corresponds to a spectral function from one sample.
In the small noise case, all reconstructed spectral functions are on one curve, which shows the robustness against the noise in the input remains for the multi-orbital case well.

\subsection{Three orbital data}
In the following, we will show a three orbits case as another demonstration.
The eigenvalues of the exact spectral function $\Lambda^\text{exact}_i(\omega)$ are defined as
\begin{equation}
\begin{aligned}
    \Lambda^\text{exact}_0(\omega) &= 0.3g(\omega; \mu=-0.5, \sigma=0.2), \\
    \Lambda^\text{exact}_1(\omega) &= 0.7g(\omega; \mu=1.3, \sigma=0.8), \\
    \Lambda^\text{exact}_2(\omega) &= 0.2g(\omega; \mu=2.2, \sigma=0.4),
\end{aligned}
\end{equation}
and the exact spectral functions are rotated as
\begin{equation}
\rho^\text{exact}(\omega) =
R^\dagger
\begin{pmatrix}
\Lambda^\text{exact}_0(\omega) & 0 & 0\\
0 & \Lambda^\text{exact}_1(\omega) & 0\\
0 & 0 & \Lambda^\text{exact}_2(\omega)
\end{pmatrix}
R,
\end{equation}
where $R = R_x(\theta_x) R_y(\theta_y) R_z(\theta_z)$ is a three dimensional rotation matrix defined as
\begin{equation}
\begin{aligned}
R_x(\theta_x) &=
\begin{pmatrix}
1 & 0 & 0 \\
0 & \cos(\theta_x) & -\sin(\theta_x) \\
0 & \sin(\theta_x) & \cos(\theta_x)
\end{pmatrix}, \\
R_y(\theta_y) &=
\begin{pmatrix}
\cos(\theta_y) & 0 & \sin(\theta_y) \\
0 & 1 & 0 \\
-\sin(\theta_y) & 0 & \cos(\theta_y)
\end{pmatrix}, \\
R_z(\theta_z) &=
\begin{pmatrix}
\cos(\theta_z) & -\sin(\theta_z) & 0 \\
\sin(\theta_z) & \cos(\theta_z) & 0 \\
0 & 0 & 1
\end{pmatrix},
\end{aligned}
\end{equation}
with $(\theta_x, \theta_y, \theta_z) = (0.3\pi, 0.2\pi, 0.4\pi)$.
As in the case of the two-orbits example,
we prepared the exact Green's function $G^\text{exact}_{ab}(\tau)$ via Eq.\ref{eq:gtau} with $\beta=100$, and generated the input data $G_{ab}(\tau)$ by adding Gaussian noises with $\sigma=10^{-6}$ to $G^\text{exact}_{ab}(\tau)$.

\begin{figure*}[tb]
\includegraphics[width=0.95\linewidth]{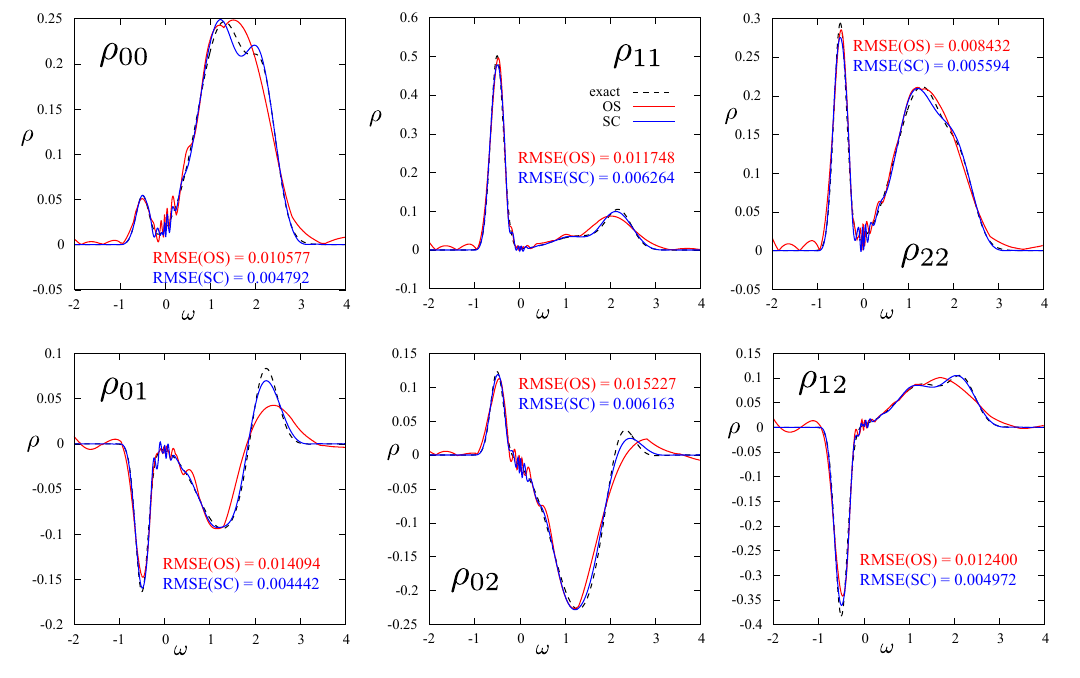}
\caption{
The reconstructed spectral functions $\rho_{ab}$ by the OS method (red curve) and the SC method (blue curve) are shown with the exact ones (black dash curve) for the three-orbital example.
Three panels on the top row show the diagonal elements and those on the bottom row do the off-diagonal ones.
Inset figures are the RMSE from the exact functions.
}
\label{fig:spectrum-3orb}
\end{figure*}

From $G_{ab}(\tau)$, we reconstructed spectral functions $\rho_{ab}(\omega)$ by the two methods.
In the SC method, we adopted $n=10$ to speed up.
Figure~\ref{fig:spectrum-3orb} shows the obtained spectral functions; the red curves are obtained by the OS method and the blue ones by the SC method.
The RMSEs from the exact spectral functions are shown as the inset figures; the RMSEs of method 1 are 2x - 3x larger than those of the SC method. In fact, we can see that the SC method gives better results than the OS method, with peaks at frequencies closer to the exact solution overall, although there is a tendency to oscillate around $\omega = 0$, as in the two-orbit case.

\section{\label{sec:summary}Summary}
In the paper, we investigate the sparse-modeling analytic continuation method from the imaginary time Green's function to the real frequency spectral function of multi-orbital systems.
The causality requires that the multi-orbital spectral function should be the semi-positive definite matrix.
To hold the semi-positive definiteness, we propose two methods of SpM-AC; the OS method is to make sure the semi-positive definiteness is at the end of the ADMM, while the SC method is to do so on every step in the self-consistent loop of the ADMM sequence.
The numerical demonstration shows that while the former is faster, the latter gives more precise results.

For applications to real materials, we sometimes need to deal with 10 or more orbital systems.
As the number of the orbits $N_\text{orb}$ increases, the computational cost of a ADMM sweep also increases as $N_\text{orb}^3$ due to diagonalization required by the SPD condition.
We also show that the reduction of the frequency points where the semi-positive definiteness condition is required achieves the speed-up with the same level of accuracy.
In this study, we adopted the uniform frequency mesh for simplicity.
On the other hand, non-uniform meshes such as the log-mesh and sparse-IR mesh are effective for reducing the computational cost because these meshes can represent the spectral functions well in a few frequency points.
For also the non-uniform mesh, it is naturally expected that
skipping the frequency points of the SPD condition can reduce the computational cost.

Our program is available from GitHub~\cite{pySpMAC}.
Dataset --- the set of $G(\tau), \rho(\omega)$, and the input parameters --- is available from ISSP Data Repository~\cite{datarepo_gitlab}.

\begin{acknowledgments}
The computation in this work has been performed on the supercomputer System B (Ohtaka) at the Supercomputer Center, the Institute for Solid State Physics, the University of Tokyo.
This study was supported by JSPS KAKENHI Grants
No. 19K03649,
No. 21H01003,
No. 21H01041,
No. 22KK0226,
No. 23H03817,
and
No. 23H04869,
Japan.
H.S. was supported by JST PRESTO Grant No. JPMJPR2012, Japan and JST FOREST Grant No. JPMJFR2232, Japan.
\end{acknowledgments}

\end{document}